\documentclass[prb,twocolumn,showpacs,preprintnumbers,amsmath,amssymb,superscriptaddress]{revtex4}
\usepackage{graphicx}
\usepackage{dcolumn}
\usepackage{bm}
\usepackage{psfrag}

\begin{document}
\title{Time-dependent Gutzwiller theory of magnetic excitations in the Hubbard model} 
\author{G. Seibold}
\affiliation{Institut f\"ur Physik, BTU Cottbus, PBox 101344, 
03013 Cottbus, Germany}
\author{F. Becca}
\affiliation{INFM-Democritos, National Simulation Centre, and SISSA I-34014 
Trieste, Italy.}
\author{P. Rubin}
\affiliation{Institute of Physics, University of Tartu,
Riia 142, 51014 Tartu, Estonia}
\author{J. Lorenzana}
\affiliation{Center for Statistical
Mechanics and Complexity, INFM, Dipartimento di Fisica,
Universit\`a di Roma La Sapienza, P. Aldo Moro 2, 00185 Roma, Italy}
\date{\today}
\begin{abstract}
We use a spin-rotational invariant Gutzwiller energy
functional to compute random-phase-approximation-like (RPA)
fluctuations on top of the Gutzwiller approximation (GA).
The method can be viewed as an extension of the previously developed 
GA+RPA approach for the charge sector
[G. Seibold and J. Lorenzana, Phys. Rev. Lett. {\bf 86}, 2605 (2001)] 
with respect to the inclusion of the magnetic excitations.
Unlike the charge case, no assumptions about the time evolution of the double
occupancy are needed in this case. Interestingly, in a spin-rotational
invariant system, we find the correct degeneracy between triplet
excitations, showing the consistency of both computations. 
Since no restrictions are imposed on the symmetry of the underlying
saddle-point solution, our approach is suitable for the evaluation of
the magnetic susceptibility and dynamical structure factor
in strongly correlated inhomogeneous systems. 
We present a detailed study of the quality of our approach
by comparing with exact diagonalization results and show its much higher
accuracy compared to the conventional Hartree-Fock+RPA
theory. In infinite dimensions, where the GA becomes exact
for the Gutzwiller variational energy, we evaluate ferromagnetic and 
antiferromagnetic instabilities from the transverse magnetic susceptibility. 
The resulting phase diagram is in complete agreement with previous
variational computations.
\end{abstract}


\maketitle

\section{Introduction}

It is now about 40 years ago that Gutzwiller proposed a variational 
wave function for correlated electronic models with a purely
local interaction, i.e., for the Hubbard-like models.~\cite{GUTZ1,GUTZ2}
The basic idea is to partially project out configurations with
doubly-occupied sites from the Fermi sea in order to optimize
the contributions from kinetic and potential energy.
As a consequence, in contrast to the conventional Hartree-Fock (HF) theory,
the Gutzwiller wave function captures correlation effects
like the band narrowing already on the variational level.
However, the exact evaluation
of the ground state energy within the Gutzwiller wave function 
is fairly difficult and
up to now has only been achieved in one and infinite dimensions.~\cite{METZNER}
In the latter case the solution is equivalent to the
so-called Gutzwiller approximation (GA) which has been 
applied to describe a variety of finite dimensional systems ranging from
the properties of normal $^3$He (see Ref.~\onlinecite{VOLLHARDT})
to the stripe phase of high-T$_c$ cuprates.~\cite{GOETZ,LOR}

The GA in its original formulation was restricted to homogeneous
paramagnetic systems and only later on generalized to arbitrary 
Slater determinants by Gebhard~\cite{GEBHARD} and, more recently, by
Attaccalite and Fabrizio.~\cite{michele}
The same energy functional
was obtained from the Kotliar-Ruckenstein (KR) slave-boson formulation
of the Hubbard model when the bosons are replaced by their
mean-values.~\cite{KR} 
Moreover, the KR slave-boson approach provides a controlled
scheme for including fluctuations beyond the mean-field solution.
Formally this has been achieved by several authors within the functional 
integral formalism.~\cite{ARRIGONI,LAVAGNA,LI,rai93,rai95} 
However, the expansion 
of the KR hopping factor $z^{SB}$ turned out to be a highly nontrivial
task, both with respect to the proper normal ordering of the bosons 
and with respect to the correct continuum limit of the functional 
integral.~\cite{ARRIGONI} 
These difficulties have severely hampered the
computation of charge fluctuations within the slave boson approach.
To our knowledge this technique has therefore only been applied 
to {\it toy models}~\cite{ARRIGONI} and to compute the
optical conductivity in the paramagnetic regime.~\cite{rai93,rai95}
The latter, however, did not lead to controlled sum rules due to the
above mentioned difficulties.~\cite{raipc}

In Refs.~\onlinecite{GOETZ1,GOETZ2} we have developed an alternative scheme 
for the computation of random-phase-approximation-like (RPA) fluctuations
beyond the GA. Our approach, labeled GA+RPA,
is based on well developed techniques in
nuclear physics~\cite{THOULESS} and RPA fluctuations are obtained
in the small oscillation limit of a time-dependent GA.
By comparing with exact diagonalization results, we have shown
that the computation of static and dynamical correlation functions
performs much better within the GA+RPA than within conventional HF+RPA
theory.~\cite{YON} Since no restrictions are imposed on the symmetry of the
saddle-point solution, the GA+RPA method is also suitable for the
investigation of strongly correlated electronically inhomogeneous systems.
Based on this formalism, two of us have recently explained the 
evolution of the optical 
conductivity with doping in high-T$_c$ cuprate compounds.~\cite{lor03}

Our previous investigations were restricted
to the evaluation of RPA fluctuations in the charge sector where
the total spin is conserved by the particle-hole 
excitations.~\cite{GOETZ1,GOETZ2}
However, in general,
one has to distinguish between longitudinal (i.e., with $\Delta S_z=0$)
and transverse spin excitations (i.e., with $\Delta S_z=\pm 1$), the latter 
involving particle-hole pairs with opposite spins.
Therefore, longitudinal excitations are optically allowed by dipole selection
rules whereas transverse excitations
can be excited by spin-carrying particles like neutrons.
For spin-rotational invariant systems,
the triplet transverse excitations with $\Delta S_z=\pm 1$ are degenerate
with the triplet longitudinal excitation with $\Delta S_z=0$ and,
therefore, it is enough to solve the problem in the longitudinal
channel. As discussed below, the solution in the transverse channel
is useful for a consistency check.   
If spin-rotational symmetry is broken,
(e.g., for ferromagnetic or spin-density-wave states) the triplet
excitations will split and one has to solve both channels to obtain 
the whole spectrum. 

The present paper is therefore devoted to the computation of
transverse magnetic excitations on top of the GA.
Various approaches have been already adopted in order to
accomplish this task.
In Ref.~\onlinecite{BM}, B\"unemann has evaluated the spin-wave excitations
in itinerant ferromagnets by determining variationally
the energy of the excited state $S_q^+|\Psi_G\rangle$, where
$|\Psi_G\rangle$ denotes the Gutzwiller wave function and $S_q^+$ is the 
spin-flip operator with momentum $q$.
Furthermore, spin excitations around paramagnetic
saddle points have been investigated in Refs.~\onlinecite{LAVAGNA,LI,FRESARD} 
within the functional integral technique
based on the spin-rotational invariant slave-boson scheme.~\cite{WH}

Our investigations below are related to these previous investigations but
differ in two important aspects.
First we will eliminate the bosonic degrees
of freedom (except for the double occupancy $D$) from the energy functional,
which thus only depends on the density matrix and the parameters ${D}$.  
Formally, this procedure defines an effective Gutzwiller Hamiltonian,
which can be expanded with respect to both charge and spin fluctuations.
As usual, both types of excitations are decoupled in case of saddle points
with collinear spin structure.
Second, the density matrix can be constructed from arbitrary 
Slater determinants, and, therefore, the method is suitable for the 
investigation of magnetic excitations in inhomogeneous systems. 
In this respect, the size limitations in numerical solutions 
are exactly the same than
for the inhomogeneous HF+RPA approach.~\cite{YON}

The paper is organized as follows.
In Sec.~\ref{section:II} we derive the GA energy functional
from the spin-rotational invariant slave-boson Hamiltonian and
show how RPA fluctuations in the charge and spin channel can
be obtained within the time-dependent Gutzwiller approach.
In particular, we focus on the magnetic excitation spectrum obtained
in this way from the Hubbard model.
Results for specific systems are presented in Sec.~\ref{section:III}.
As a first example, we consider in Sec.~\ref{section:IIIA} 
the two-site Hubbard model, where the analytical solution is available
for comparison. Since at small $U$ the mean field  
ground state is spin-rotationally invariant, the expected degeneracy
between longitudinal and transverse spin excitation allows us to check
the consistency among charge and magnetic channel computations.  
Then, in Sec.~\ref{section:IIIB}, the method is applied to a homogeneous and
paramagnetic GA solution, where it turns out that
the evaluation of transverse magnetic susceptibilities
is greatly simplified as compared to previous approaches.
In particular, we evaluate the ferromagnetic and 
antiferromagnetic
instability lines for an infinite-dimensional hypercubic system,
and demonstrate the exact agreement with variational results.
Section~\ref{section:IIIC} is devoted to a comparison of the GA+RPA 
magnetic excitation  
spectra with exact diagonalization and HF+RPA results respectively.
Concluding remarks appear in Sec.~\ref{section:IV}.

\section{Formalism}\label{section:II}

\subsection{Spin-rotational invariant GA}

The starting point is the one-band Hubbard model:
\begin{equation}\label{HM}
H= \sum_{i,j,\sigma} t_{ij} c_{i,\sigma}^{\dagger}c_{j,\sigma} + U\sum_{i}
n_{i,\uparrow}n_{i,\downarrow},
\end{equation}
where $c_{i,\sigma}$ ($c^\dagger_{i,\sigma}$) destroys (creates) an electron 
with spin $\sigma$ at site
$i$, and $n_{i,\sigma}=c_{i,\sigma}^{\dagger}c_{i,\sigma}$. $U$ is the
on-site Hubbard repulsion and $t_{ij}$ denotes the hopping parameter between
sites $i$ and $j$. 
Our investigations are based on the spin-rotational invariant form of the
slave-boson approach introduced by KR.~\cite{WH} 
Within this formalism one introduces auxiliary bosons 
$e_{i}$ ($e_{i}^\dagger$) and $d_{i}$ ($d_{i}^\dagger$)
which represent the annihilation (creation) of
empty and doubly-occupied sites, respectively.
In addition, the single occupied states are represented by
two particles, a spin-1/2 fermion and a boson $p$ which  
can have either spin $S=0$ or $S=1$ in such a way that the 
combination has spin-1/2. The four $p$ states 
(a singlet and a triplet) are combinations of the elements  
$p_{i,\sigma\sigma'}$ of a $2\times2$ matrix  ${\bf p}_i$.
In the saddle-point approximation all boson operators are treated 
as numbers and the matrix ${\bf p}_i$ can be parametrized as:
\begin{equation}\label{PI}
{\bf p}_i = \left(\begin{array}{cc}
p_{i,\uparrow} & \frac{1}{\sqrt{2}}p_i \exp{(-i\phi_i)} \\
\frac{1}{\sqrt{2}}p_i \exp{(+i\phi_i)} & p_{i,\downarrow} \end{array}\right),
\end{equation}
with $p_i$, $p_{i\sigma}$, and $\phi_i$ real.

Besides the completeness condition
\begin{equation}\label{CONST1}
e_{i}^{2}+ \mbox{tr} ({\bf p}_i^* {\bf p}_i)+D_{i}=1,
\end{equation}
the boson fields are constrained by the following relations
\begin{equation}\label{CONST2}
\mbox{tr}({\bf \tau}_{\mu} {\bf p}_i^{*} {\bf p}_i)+2 \delta_{\mu,0} D_{i}
=\sum_{\sigma,\sigma'}({\bf \tau}_{\mu})_{\sigma,\sigma'}  
\rho_{ii}^{\sigma,\sigma'},
\end{equation}
where, in general, $\rho_{ij}^{\sigma,\sigma'}
\equiv \langle c^\dagger_{i,\sigma}c_{j,\sigma'} \rangle$
denotes the density matrix,
${\bf \tau_{\mu}}$ are the Pauli matrices (including
${\bf \tau}_{0} \equiv {\bf 1}$), and  $D_i\equiv d_i^2$.
 
After rewriting the Hamiltonian~(\ref{HM}) in terms
of fermion and boson operators,~\cite{WH} 
we can construct a {\it spin-rotational invariant Gutzwiller functional} 
by eliminating the boson fields 
except for $D_i$
via the constraints, Eqs.~(\ref{CONST1}) and~(\ref{CONST2}). As a result,
one obtains:
\begin{equation}\label{EGA}
E^{GA}= \sum_{i,j,\sigma,\sigma_1,\sigma_2}
t_{ij} z_{i,\sigma_1,\sigma}
z_{j,\sigma,\sigma_2} \rho_{ij}^{\sigma_1,\sigma_2} 
+ U\sum_{i}D_{i},
\end{equation}
where the matrix ${\bf z}_i$ reads as:
\begin{equation}\label{zmat}
{\bf z}_i=\left( \begin{array}{cc} 
{z_i}^+\cos^2\frac{\Phi}{2}+{z_i}^-\sin^2\frac{\Phi}{2} & 
\frac{S_i^-}{S_i^z}[{z_i}^+-{z_i}^-]\cos\Phi \\ 
\frac{S_i^+}{S_i^z}[{z_i}^+-{z_i}^-]\cos\Phi & 
{z_i}^+\sin^2\frac{\Phi}{2}+{z_i}^-\cos^2\frac{\Phi}{2}
\end{array} \right),
\end{equation}
with
\begin{eqnarray}
\tan^2\Phi&=&\frac{S_i^+S_i^-}{(S_i^z)^2},\\
{z_i}^\pm &=& \frac{\sqrt{1-\rho_{ii}+D_i}\lambda_i^\pm
+\lambda_i^{\mp}\sqrt{D_i}}{\sqrt{\left(1-D_i-(\lambda_i^\pm)^2\right)
\left(\rho_{ii}-D_i-(\lambda_i^\mp)^2\right)}},\\
(\lambda_i^\pm)^2 &=& \rho_{ii}/2 - D_i \pm S_i^z\sqrt{1+\tan^2\Phi},
\end{eqnarray}
and for clarity spin expectation values are denoted by
$S_i^+=\rho_{ii}^{\uparrow,\downarrow}$, 
$S_i^-=\rho_{ii}^{\downarrow,\uparrow}$, 
$S_i^z=(\rho_{ii}^{\uparrow,\uparrow}-\rho_{ii}^{\downarrow,\downarrow})/2$,
and 
$\rho_{ii}=\rho_{ii}^{\uparrow,\uparrow}+\rho_{ii}^{\downarrow,\downarrow}$.
Note that in the limit $S_i^\pm=0$, where the matrix ${\bf z}_i$ is
diagonal, one recovers the standard Gutzwiller energy functional
as derived by Gebhard~\cite{GEBHARD} or 
KR.~\cite{KR} Furthermore, it has been shown that
the spin-rotational invariant slave-boson scheme can be derived from
the KR (or alternatively Gebhard's) energy functional when the spin rotation
is applied to the underlying Slater determinant.~\cite{MICNAS}
Therefore, Eq.~(\ref{EGA}) can be viewed as the more general GA-like
energy functional for a Hubbard Hamiltonian. 

In order to obtain the stationary solution of Eq.~(\ref{EGA}) 
one has to minimize $E^{GA}$
with respect to the double occupancy parameters $D$ and the density
matrix $\rho$. The latter variation has to be constrained to the 
subspace of Slater determinants by imposing the condition
\begin{equation}\label{SDC}
\rho^2=\rho,
\end{equation}
which is equivalent to the
diagonalization of the electronic problem supplemented by the
variation with respect to $D$ only.
A detailed description of the corresponding formalism can be found
in Ref.~\onlinecite{SSH}.

Regarding the stationary solutions, we will restrict to
Slater determinants which are diagonal in spin space, i.e.,
$\rho_{ij}^{\sigma,\sigma'(0)}=
\rho_{ij}^{\sigma,\sigma(0)}\delta_{\sigma,\sigma'}$.
Thus we do not consider spin canted solutions~\cite{goe98} which would mix
charge and spin excitations. Therefore, the diagonalized density matrices have
eigenvalue 1 below the Fermi level ($\equiv$ hole states: h) 
and zero above ($\equiv$ particle
states: p) and consequently are also diagonal in spin space:
\begin{eqnarray}
\rho^{(0)}_{h\sigma,h\sigma} &=& 1 \label{P1},\\
\rho^{(0)}_{p\sigma,p\sigma} &=& 0\label{P2}.
\end{eqnarray}
Within this notation we can formally write the GA energy as
\begin{equation}
E^{GA}=\sum_{k\sigma}\epsilon_k \rho^{(0)}_{k\sigma,k\sigma}+U\sum_i D_i,
\end{equation}
where $k=p,h$ labels particle and hole states and $\epsilon_{k}$ 
are the corresponding one-particle energies.

\subsection{Calculation of RPA fluctuations around general GA saddle points
and magnetic excitations} 

The energy functional Eq.~(\ref{EGA}) is a convenient starting point
for the calculation of charge and spin excitations
on top of general GA wave functions.
In Refs.~\onlinecite{GOETZ1,GOETZ2}, we have already given a detailed 
derivation of the GA+RPA formalism in the charge sector,
which, in the following, we extend to include the spin fluctuations.

We thus study the response of the system to an external
time-dependent perturbation
\begin{eqnarray}\label{eq:fdt}
F(t)&=&
\sum_{i,j,\sigma,\sigma'} [f_{ij,\sigma\sigma'}(t)
c_{i,\sigma}^{\dagger} c_{j,\sigma'} + H.c.], \\
f_{ij,\sigma\sigma'}(t)&=&f_{ij,\sigma\sigma'}(0)e^{-i\omega t},
\end{eqnarray}
which induces small amplitude oscillations
of $D$ and $\rho$ around the GA saddle point:
\begin{eqnarray}
D &=& D^{(0)} + \delta D(t), \\
\rho &=& \rho^{(0)} + \delta \rho(t).
\end{eqnarray}
Correspondingly, we have to expand the energy functional Eq.~(\ref{EGA})
around the stationary solution up to second order in the density- and 
double-occupancy deviations.
Due to the fact that we restrict to collinear saddle-point solutions,
the charge and spin sectors in the expansion are decoupled and one 
obtains 
\begin{equation} \label{E2}
E[\rho,D]=E_0+\mbox{tr}\{h^0 \delta{\rho}\} + \delta E^{charge} 
+ \delta E^{spin}, 
\end{equation}
where the subscript 0 indicates quantities evaluated in the stationary state 
and we have introduced the Gutzwiller Hamiltonian:~\cite{RING,BLAIZOT}
\begin{equation}
  \label{eq:hgw}
  h_{ij}^{\sigma,\sigma'}[\rho,D]=\frac{\partial E^{GA} } 
{\partial \rho_{ji}^{\sigma',\sigma}}\delta_{\sigma,\sigma'}.
\end{equation}
$\delta E^{charge}$ contains the expansion with respect to the double-occupancy
parameters and the part of the density matrix, which is diagonal in the 
spin indices.
This part of the RPA problem has already been studied in detail in
Refs.~\onlinecite{GOETZ1,GOETZ2}, where it was shown that
the $\delta D$ fluctuations can be eliminated 
by assuming that they adjust instantaneously to the evolution
of the density matrix (antiadiabaticity condition).

The spin part of the expansion reads:
\begin{eqnarray}
&&\delta E^{spin}= \label{espin}
\sum_{i,j,\sigma} t_{ij} \rho_{ij}^{(0)\sigma,\sigma}
\lbrack z^{0}_{i,\sigma,\sigma} \delta_2 z_{j,\sigma,\sigma}
+ z^{0}_{j,\sigma,\sigma} \delta_2 z_{i,\sigma,\sigma}\rbrack
 \nonumber \\
&&+\sum_{i,j,\sigma} t_{ij} z^{0}_{i,\sigma,\sigma}  \lbrack 
\delta_1 z_{j,\sigma,-\sigma} \delta\rho_{ij}^{\sigma,-\sigma}
+ \delta_1 z_{j,-\sigma,\sigma}\delta\rho_{ji}^{-\sigma,\sigma} 
\rbrack \nonumber \\
&&+\sum_{i,j,\sigma} t_{ij} \rho_{ij}^{(0)\sigma,\sigma}
\delta_1 z_{i,\sigma,-\sigma} \delta_1 z_{j,-\sigma,\sigma},
\end{eqnarray}
with the following abbreviations for 
the quadratic parts of the z-factor expansion
\begin{eqnarray} \label{z1}
\delta_1 z_{i,\sigma,-\sigma} &=& 
\frac{\partial z_{i,\sigma,-\sigma}}{\partial \rho_{ii}^{-\sigma,\sigma}}
\delta \rho_{ii}^{-\sigma,\sigma}, \\
\delta_2 z_{i,\sigma,\sigma} &=& 
\frac{\partial^2 z_{i,\sigma,\sigma}}
{\partial \rho_{ii}^{\sigma,-\sigma_2}\partial \rho_{ii}^{-\sigma,\sigma}}
\delta \rho_{ii}^{\sigma,-\sigma}\delta \rho_{ii}^{-\sigma,\sigma}.
\label{z2}
\end{eqnarray}
The explicit results for the derivatives are given in appendix~\ref{APA}. 
It is interesting to observe that, in contrast to the charge excitations,
the evaluation of the magnetic excitations can be performed
without any adjustment of $\delta D$ to $\delta \rho$, i.e., without
any assumption on the time evolution of $D$.
Only in the case of non-collinear saddle points one would have a coupling 
between spin and charge fluctuations
and, therefore, the necessity to invoke the antiadiabaticity condition
to eliminate the $\delta D$ deviations. 

The density fluctuations $\delta\rho$ in the expansion Eq.~(\ref{E2}) 
are restricted to the subspace of Slater determinants, i.e.,
they have to obey the constraint Eq.~(\ref{SDC}).
One can therefore divide $\delta\rho$ 
into the particle (p) and hole (h) sectors 
using the property of the density matrices Eqs.~(\ref{P1}) and~(\ref{P2}):
\begin{eqnarray}
\{\delta\rho^{hp}_{\sigma\sigma'}\}
&\equiv&\rho^{(0)}_{\sigma\sigma}\delta\rho(1-\rho^{(0)}_{\sigma'\sigma'}), 
\label{eq:fl1}\\ 
 \{\delta\rho^{ph}_{\sigma\sigma'}\}&\equiv&(1-\rho^{(0)}_{\sigma\sigma})
\delta\rho\rho^{(0)}_{\sigma'\sigma'},
\label{eq:fl2}\\
\{\delta\rho^{hh'}_{\sigma\sigma'}\}&\equiv&\rho^{(0)}_{\sigma\sigma}
\delta\rho(1-\rho^{(0)}_{\sigma'\sigma'}), 
\label{eq:fl3}\\
\{\delta\rho^{pp'}_{\sigma\sigma'}\}&\equiv&(1-\rho^{(0)}_{\sigma\sigma})
\delta\rho\rho^{(0)}_{\sigma'\sigma'}.
\label{eq:fl4}
\end{eqnarray}
where by $\{\delta\rho^{hp}_{\sigma\sigma'}\}$ we mean a matrix 
whose non-zero generic elements are of the form 
$\delta\rho^{hp}_{\sigma\sigma'}$.
Moreover, one can show [see Eqs.~(34)-(36) in Ref.~\onlinecite{GOETZ2}] that
the $pp$ and $hh$ density projections yield a quadratic contribution
in the $ph$ and $hp$ matrix elements in the small amplitude approximation: 
\begin{eqnarray}
\delta\rho^{hh}_{\sigma\sigma'}&\approx&
-\sum_{p\sigma''}\delta\rho^{hp}_{\sigma\sigma''}\delta\rho^{ph}_{\sigma''\sigma'}, 
\label{hhpp1}\\
\delta\rho^{pp}_{\sigma\sigma'}&\approx&\sum_{h\sigma''}
\delta\rho^{ph}_{\sigma\sigma''}\delta\rho^{hp}_{\sigma''\sigma'}. 
\label{hhpp2}
\end{eqnarray}
Hence, although the Gutzwiller Hamiltonian Eq.~(\ref{eq:hgw}) is diagonal
in spin space, it turns out that the term
$\mbox{tr}(h^0\delta\rho)=
\sum_{\mu} \epsilon_{\mu} \rho_{\mu\mu}$ in Eq.~(\ref{E2}) (which is first
order in the $pp$ and $hh$ density projections)
yields a quadratic contribution in the $ph$ and $hp$ matrix elements:
\begin{eqnarray}
\mbox{tr}(h^0\delta\rho)&=&\sum_{p\sigma} \epsilon_{p}
\delta\rho^{pp}_{\sigma\sigma}
                      +\sum_{h\sigma} \epsilon_{h}
\delta\rho^{hh}_{\sigma\sigma}, \nonumber \\ \label{trh}
&=& \sum_{ph\sigma\sigma'}(\epsilon_p - \epsilon_h)
\delta\rho^{ph}_{\sigma\sigma'}\delta\rho^{hp}_{\sigma'\sigma}.
\end{eqnarray}
The fluctuations which are diagonal in the spin indices
($\delta\rho^{ph}_{\sigma\sigma}$ and $\delta\rho^{hp}_{\sigma\sigma}$)
contribute to the expansion in the charge channel,~\cite{GOETZ2}
whereas the non-diagonal elements describe the
zero-order (non-interacting) spin-flip excitations of the saddle-point
Slater determinant.

Thus, up to second order in the particle-hole (spin) density fluctuations, 
one obtains
for the energy expansion:
\begin{equation}\label{phe}
\delta E^{spin}=\frac{1}{2}
(\delta\rho^{hp} ,\delta\rho^{ph})
  \left( \begin{array}{cc}
A & B \\
B^{*} & A^{*} \end{array}\right)
\left(\begin{array}{c}
 \delta\rho^{p'h'} \\
 \delta\rho^{h'p'}
\end{array}\right),
\end{equation}
where the explicit expressions for the RPA matrices $A$ and $B$ are given in the
appendix~\ref{APB}. Note that the shorthand notation in Eq.~(\ref{phe}) and
below implies that p- and h-states have opposite spin, i.e.,
$\delta\rho^{ph}$ represent the joint set of elements of types
$\delta\rho^{ph}_{\uparrow\downarrow}$
and $\delta\rho^{ph}_{\downarrow\uparrow}$. 
Following Ref.~\onlinecite{GOETZ2}, we can now evaluate the response
function corresponding to the perturbation Eq.~(\ref{eq:fdt}).
In case of non-diagonal perturbations (as the coupling to a current),
one has to define an associated Gutzwiller operator which contains
the GA hopping matrices ${\bf z}_i$.
However, in the spin channel the most relevant perturbations couple an
external field locally to some spin operator. The field
$f_{ij,\sigma\sigma'}=f_{ii,\sigma\sigma'}\delta_{ij}$
is therefore diagonal in the site representation and remains
unchanged within the GA.
Upon transforming the perturbation to the particle-hole representation
one can derive the following linear response equation:
\begin{equation}
  \label{eq:rpa}
 \left\{ \left( \begin{array}{cc}
A & B \\
B^{*} & A^{*} \end{array} \right)-
\hbar \omega
\left(\begin{array}{cc}
1 & 0 \\
0 & -1
\end{array}\right)\right\}\left(\begin{array}{c}
\delta\rho^{ph}\\
\delta\rho^{hp}
\end{array}\right) =-
\left(
\begin{array}{c}
 f_{ph}\\
 f_{hp}
\end{array}\right).
\end{equation}
The inversion of Eq.~(\ref{eq:rpa}) yields a 
linear relation between the external field
and the change in the density:
\begin{equation}\label{eq:lr}
\delta\rho=R(\omega) f,
\end{equation}
and defines the linear response function $R(\omega)$
which in the  Lehmann
representation reads as:
\begin{equation}\label{eq:leh}
R(\omega)_{ph,p'h'}=\sum_{n>0}\left\lbrack
\frac{X_{ph}^n X_{p'h'}^{n*}}{\omega-\Omega_{n}+i\epsilon} -
\frac{Y_{p'h'}^n Y_{ph}^{n*}}{\omega+\Omega_{n}+i\epsilon}\right\rbrack,
\end{equation}
where we have introduced the eigenvectors of the RPA matrix:
\begin{eqnarray}
\langle 0 | a^{\dagger}_{h} a_p |n \rangle&\equiv& X_{ph}^{n}, \label{eq:xy1}\\
\langle 0 | a^{\dagger}_p a_h |n \rangle&\equiv&Y_{hp}^{n}. \label{eq:xy2}
\end{eqnarray}
and $|n\rangle$ denote the unprojected (i.e., without Gutzwiller correlations)
excited states of the RPA problem.

\section{Results}\label{section:III}

The RPA formalism derived in the previous section constitutes
a convenient starting point for the calculation of spin excitations
on top of the GA. One of the advantages of the
present approach is that it is suitable for general Slater determinants,
i.e., without any restriction on translational and (longitudinal)
spin symmetries. The system sizes which can be treated  are
the same than for the traditional HF+RPA approximation.~\cite{YON}
However, also for homogeneous and paramagnetic saddle points the GA
based RPA approach provides a convenient method for the evaluation of
spin fluctuations. Our method is solely based on the
expansion of the density matrix in terms of particle-hole fluctuations
and does not involve other degrees of freedom as in the
related functional integral slave-boson
scheme.~\cite{LAVAGNA,LI,FRESARD}
First, this advantage is demonstrated 
for a two-site Hubbard model which is also a convenient {\it toy model}
for the RPA formalism derived in the previous
section. However, the GA+RPA approach can also be applied within
the more conventional Green's function technique which is used in 
Sec.~\ref{section:IIIB}
to evaluate spin susceptibilities for a homogeneous and paramagnetic
hypercubic lattice in infinite dimensions. In this case 
the GA becomes exact for the energy functional within the Gutzwiller 
wave function, and we recover the magnetic instability lines determined 
previously by Fazekas and collaborators.~\cite{FAZEKAS} 
The remainder of this section is then devoted to a detailed
analysis of the quality of our approach by comparing with HF+RPA and 
exact results for small clusters, where the exact solution is known by
exact diagonalization techniques.

\subsection{Two-site Hubbard model}\label{section:IIIA}

As a first example, we consider the two-site Hubbard model at half filling
which can be
solved exactly and can be studied analytically with both the GA+RPA
and HF+RPA approximations. On general grounds a mean-field (or
time-dependent mean-field) approach is
expected to improve as the dimensionality of the space increases, and,
therefore, this zero-dimensional problem is the worst case
and may give an estimation of the maximum error which can be expected for
these mean-field approaches. 

The exact ground-state energy is given by:
\begin{equation}
E_0=\frac{1}{2}\left[ U-\sqrt{U^2+16t^2}\right],
\end{equation}
and the corresponding eigenfunction reads as
\begin{eqnarray}
|\Psi_0\rangle&=&\alpha \frac{1}{\sqrt{2}}\left[|\uparrow_1\downarrow_2\rangle
- |\uparrow_2\downarrow_1\rangle \right] \nonumber \\
&+&\beta \frac{1}{\sqrt{2}}\left[|\uparrow_2\downarrow_2\rangle
- |\uparrow_1\downarrow_1\rangle \right],\\
\alpha^2 &=& \frac{4t^2}{E_0^2+4t^2}\,\,\,;\,\,\, \alpha^2+\beta^2=1.
\end{eqnarray}
Moreover, only the antisymmetric combination of the spin-flip operators
\begin{equation}
S_{\pi}^\pm = \frac{1}{\sqrt{2}}\left[ S_1^\pm - S_2^\pm\right]
\end{equation}
induces a transition to a state with energy $E=0$,
so that the excitation energy is given by
$\omega_{spin}^{ex}=-E_0$.

Notice that the exact solution does not display a phase transition but 
remains paramagnetic (and analytic) in $U/t$.
On the other hand, in the HF theory, one finds a paramagnetic solution 
below $U_{crit}^{HF}/t=2$ and a N\'eel-type ordered solution for 
$U>U_{crit}^{HF}$.
The latter is clearly non-physical and related to the mentioned
limitation of mean-field in a low-dimensional system.   
In the GA approximation, the electronic correlations are approximated in a 
better way and the range of the paramagnetic solution is extended, 
giving rise to:
$$U_{crit}^{GA}/t =  8(\sqrt{2}-1)\approx 3.31$$

Since the analytic expressions for the symmetry-broken regime
become quite lengthy,
we restrict the derivation below to the paramagnetic case,
where the expansion of the energy functional is given by
\begin{eqnarray}
\delta E^{spin}&=&\frac{4U_s}{N} \sum_{q=0,\pi} 
\delta S^+_q \delta S_{-q}^-,
\nonumber \\
U^s &=& -u \frac{(2+u)(1-u)}{1+u} t\,\,\,;\,\,\,u=U/(8t). \nonumber
\end{eqnarray}
Note that within the HF+RPA approximation, we have $U^s=-U/4$.

The RPA matrices read as:
\begin{equation}
A=\left( \begin{array}{cc} \Delta E + 2U^s & 0 \\ 0 & \Delta E + 2U^s
\end{array} \right) \,\,\, ; \,\,\,
B=\left( \begin{array}{cc} 0 & 2U^s \\ 2U^s & 0
\end{array} \right), \nonumber
\end{equation}
with the one-particle excitation energies $\Delta E = 2t(1-u^2)$.

The diagonalization of the eigenvalue problem yields two degenerate
excitation energies:
\begin{equation}
\omega_{\lambda=1,2}^2 \equiv \Omega^2 = \Delta E \lbrack \Delta E + 4 U^s \rbrack.
\end{equation}
Since the ground state is a singlet, these energies in the spin channel 
coincide with the longitudinal magnetic excitations computed 
in the charge channel (see Ref.~\onlinecite{GOETZ2}). Correspondingly,
one has three triplet excitations in total, with $\Delta S_z=-1,0,1$.
This indicates that the spin-rotation invariance
is correctly implemented in our approach, a fact that is far from being 
trivial.
It is worth noting that in the charge channel
an extra assumption was needed, namely the antiadiabatic   
adjustment of the double occupancy to the time evolution of the
density matrix, which was not necessary for the present calculation.
Therefore, the fact that the spin-rotation symmetry is preserved 
among both independent computations can be used as a justification 
{\it a posteriori} of the previous assumption.

\begin{figure}[tbp]
\psfrag{ylabel}{\tiny $\int Im\chi^{+-}\!d\omega$}
\includegraphics[width=7cm,clip=true]{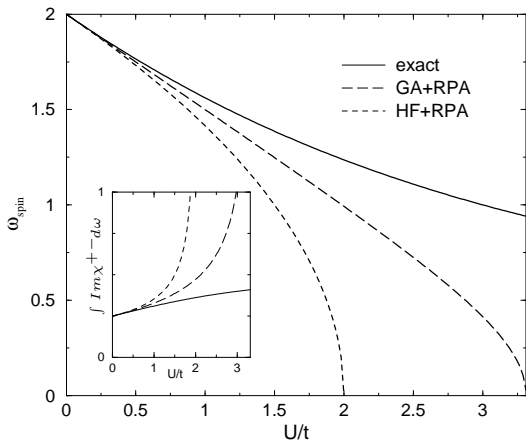}
\caption{ Spin excitation energies for a two-site system computed with GA+RPA,
HF+RPA and the exact result. In the inset:
$\int^\infty_{0}Im \chi^{+-}_{\pi}(\omega) d\omega$ for 
the same methods.}
\label{2site}
\end{figure}

Fig.~\ref{2site} shows the magnetic excitation energies for
the GA+RPA, the HF+RPA and the exact solution. 
Although in the exact solution there is no transition at finite $U$ and
the magnetic excitation energies are always finite, in the mean-field 
approximations there is a transition to a symmetry-broken 
spin-density-wave (SDW) state, marked by the vanishing of $\Omega$ 
at the respective critical values $U_{crit}^{GA/HF}$.

The dynamical transverse susceptibility is given by
\begin{equation}
\chi^{+-}_{q}(\omega)=\frac{1}{2}\sum_{\lambda} 
\frac{\langle RPA |S_q^+|\lambda\rangle
\langle \lambda|S_q^{-}|RPA\rangle}{\omega - \omega_{\lambda}+i\epsilon},
\end{equation}
which is only different from zero for $q=\pi$ since
\begin{equation}
\sum_{\lambda} |\langle RPA|S_q^{\pm}|\lambda \rangle|^2 = 
\frac{\Delta E}{2 \Omega} \delta_{q,\pi}.
\end{equation}
In the inset of Fig.~\ref{2site}, we compare the integrated susceptibility
$\int^\infty_{0} Im \chi^{+-}_{\pi}(\omega) d \omega$
for the HF+RPA, the GA+RPA and the exact result as a function of $U/t$.
The transition to the SDW state in the GA and HF approximation 
is characterized by diverging spin-spin correlations at $q=\pi$, which 
is due to the vanishing of $\Omega$ at $U_{crit}^{GA/HF}$.

The error in both the excitation energies and the integrated susceptibility 
is more than halved in the GA+RPA with respect to the HF+RPA. 
This can be traced back to the fact that, within the GA, we have a better 
ground state with a delayed instability as a function of $U/t$. 
Close to the instability one has the worst performance in a mean-field plus RPA
approach since in this  region the harmonic approximation to the energy 
functional breaks down.  

\subsection{Paramagnetic regime in infinite dimensions}\label{section:IIIB}
As a further application and to get more insight into our approximation, 
we apply the GA+RPA method to an infinite-dimensional hypercubic lattice,
where the performance is expected to be the best. 
We consider a partially filled system with density $n=1-\delta$.

The on-site elements of the density matrix for a 
paramagnetic saddle-point solution are given by
$\rho_{ii}^{\sigma,\sigma'}=\frac{n}{2}\delta_{\sigma\sigma'}$,
so that the matrix ${\bf z}_i$ of Eq.~(\ref{zmat}) reads as
\begin{equation}\label{zmat2}
{\bf z}_i=\left( \begin{array}{cc} 
z_0 & 0 \\
0 & z_0 \end{array} \right),
\end{equation}
where, by using the notation introduced by Vollhardt in 
Ref.~\onlinecite{VOLLHARDT}, we have:
\begin{eqnarray}
z_0 &=& \sqrt{\frac{2x^2-x^4-\delta^2}{1-\delta^2}}\label{z0}, \\
x &=& \sqrt{1-n+D}+\sqrt{D}. \label{xd}
\end{eqnarray}
For the Gutzwiller approximated energy one obtains 
\begin{eqnarray}\label{EGAK}
E^{GA}&=&N z_0^2e_0 + N U D, \\
z_0^2e_0&=&\frac{1}{N}\sum_{k\sigma} \varepsilon_k \rho^{\sigma,\sigma}_{kk},
\end{eqnarray}
where $e_0$ denotes the energy per site of the 
non-interacting system, $\varepsilon_k$ is the electronic dispersion 
corresponding to the Gutzwiller Hamiltonian~(\ref{eq:hgw}) and $N$ is
the number of sites.
The minimization of Eq.~(\ref{EGAK}) yields
\begin{equation}
\frac{x^4(1-x^2)}{x^4-\delta^2}=(1-\delta^2)
\frac{U}{8|e_0|} \equiv u,
\end{equation}
which, by using Eq.~(\ref{xd}), determines the double-occupancy parameter $D$.

The energy expansion Eq.~(\ref{espin}) in the momentum space is given by:
\begin{eqnarray}
\delta E^{spin}&=&\frac{1}{N} \sum N_q \delta S^+_q \delta S_{-q}^- 
\nonumber \\
&+& \frac{1}{N} \frac{z'}{z_0} \sum_q \left(\delta T_q^+\delta S_{-q}^-
+\delta S_q^+ \delta T_{-q}^-\right), \label{eskspace}
\end{eqnarray}
with the following definitions:
\begin{eqnarray}
N_q &=& 2e_0z_0 z''+ 
\left(\frac{z'}{z_0}\right)^2 \frac{1}{N}
\sum_{k\sigma} \varepsilon_{k+q}\rho_{kk}^{\sigma,\sigma}, \\
\delta S^\sigma_q &=& \sum_k \delta \rho_{k+q,k}^{\sigma,-\sigma}, \\
\delta T^\sigma_q &=& \sum_k \left(\varepsilon_{k \pm q}+\varepsilon_k\right) 
\delta \rho_{k+q,k}^{\sigma,-\sigma},
\end{eqnarray}
and the derivatives $z'$ and $z''$ are given in appendix~\ref{APA}.

Within the RPA approach presented in Sec.~\ref{section:II}, one always 
computes all excitation energies, which constitutes a suitable procedure for
the solutions on finite clusters. In infinite system it is
usually more convenient to treat the RPA problem in terms of a 
conventional Dyson approach. Therefore, we use the well known
equivalence between both formulations~\cite{YON} to set up a 
Dyson equation. 
The interaction kernel which enters the S-matrix
in the Green's function description can be formally  obtained from
Eq.~(\ref{eskspace}) by substituting the density matrix fluctuations
by the corresponding operator expressions, for instance:
\begin{eqnarray*}
\delta S^+_q \to S^+_q &=& \sum_k c^\dagger_{k+q,\uparrow}c_{k,\downarrow}, \\
\delta T^+_q \to T^+_q &=& \sum_k (\varepsilon_{k+q}+\varepsilon_k)
c^\dagger_{k+q,\uparrow}c_{k,\downarrow}.
\end{eqnarray*}

Since the energy expansion Eq.~(\ref{eskspace}) is a quadratic form in
$\delta S_q^\pm$ and $\delta T_q^\pm$ it is useful to define
the following matrix for the bare time-ordered correlation functions 
\begin{equation}\label{chimat}
{\bf \chi}_q^0(t)=\frac{i}{N} \left( \begin{array}{cc} 
\langle {\cal T} S^+_q(t) S_{-q}^-(0)\rangle_0 & \langle {\cal T} S^+_q(t) T_{-q}^-(0) \rangle_0 \\
\langle {\cal T} T^+_q(t) S_{-q}^-(0)\rangle_0 & \langle {\cal T} T^+_q(t) T_{-q}^-(0) \rangle_0 
\end{array} \right),
\end{equation}
where, the notation $\langle\dots\rangle_0$ indicates that the correlation 
functions are calculated from the excitation spectrum of the Gutzwiller 
Hamiltonian Eq.~(\ref{eq:hgw}) and~(\ref{trh}) and as a function of frequency 
one obtains:
\begin{widetext}
\begin{equation}\label{sus}
{\bf \chi}_q^0(\omega)=-\frac{1}{N} \sum_k\left( \begin{array}{cc} 
1 & \varepsilon_k + \varepsilon_{k+q}\\
\varepsilon_k + \varepsilon_{k+q} & (\varepsilon_k + \varepsilon_{k+q})^2 \end{array} \right) 
\left[\frac{n_{k+q,\uparrow}(1-n_{k,\downarrow})}
{\omega+\varepsilon_{k+q}-\varepsilon_k+i\delta} - 
\frac{n_{k,\downarrow}(1-n_{k+q,\uparrow})}
{\omega+\varepsilon_{k+q}-\varepsilon_k-i\delta}\right] .
\end{equation}
\end{widetext}
The RPA series for the spin excitations then corresponds to the
following Dyson equation:
\begin{equation}\label{eqrpa}
{\bf \chi}_q(\omega) = {\bf \chi}_q^0(\omega) - 
{\bf \chi}_q^0(\omega) {\bf M}_q {\bf \chi}_q(\omega),
\end{equation}
with the interaction kernel:
\begin{equation}\label{mmat}
{\bf M}_q=\left( \begin{array}{cc} 
N_q & \frac{z'}{z_0} \\
\frac{z'}{z_0} & 0 \end{array} \right).
\end{equation}

\begin{figure}[tbp]
\includegraphics[width=7cm,clip=true]{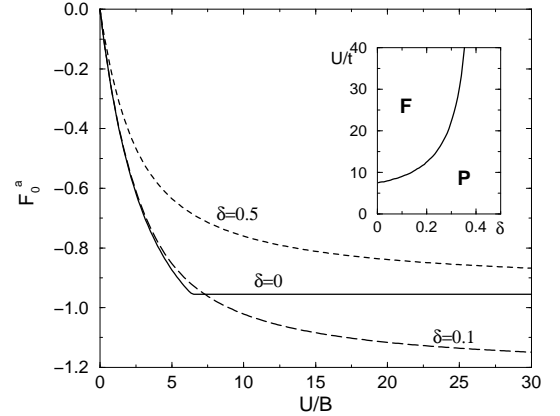}
\caption{Landau parameter $F_0^a$ as function of $U/B$ for an 
infinite-dimensional hypercubic lattice. The inset shows the 
paramagnetic-ferromagnetic instability line.}
\label{figfla}
\end{figure}

As a check of the consistency of our approach, we determine
the paramagnetic-ferromagnetic and
paramagnetic-antiferromagnetic phase boundaries.
This can be compared with previous results within the GA
obtained by evaluating the vanishing of the corresponding order
parameter.
In case of the ferromagnetic instability we have to analyze
the limit $\lim_{q\to 0}{\bf \chi}_q(\omega=0)$
so that the susceptibility matrix simplifies to
\begin{equation}\label{mmat2}
{\bf \chi}_0^0(\omega=0)=N(\varepsilon_F)\left( \begin{array}{cc} 
1 & 2 \varepsilon_F \\
2 \varepsilon_F & 4 \varepsilon_F^2 \end{array} \right),
\end{equation}
where $N(\varepsilon_F)$ denotes the density of states at the
Fermi level $\varepsilon_F$.
The inversion of Eq.~(\ref{eqrpa}) yields as a condition for the
existence of a pole at $\omega=0$ and $q=0$:
\begin{equation} 
{\rm Det} \left[{\bf 1}+ {\bf \chi}_0^0(\omega=0) {\bf M}_0\right] 
\equiv 1+F_0^a =0,
\end{equation}
with the Landau parameter $F_0^a$
\begin{equation}
F_0^a = N(\varepsilon_F)\left[e_0(2z_0z'' + z'^2) + 4\varepsilon_F\frac{z'}{z_0}
\right] .
\end{equation}
In the half-filled case ($\delta=0$) and a symmetric density of states
($\varepsilon_F=0$) this expression naturally coincides with
Vollhardt's result [see Eq.(61) in Ref.~\onlinecite{VOLLHARDT}].
Fig.~\ref{figfla} displays $F_0^a$ for a Gaussian density of states
\begin{equation}
N(\omega)=\frac{1}{\sqrt{2\pi}B}\exp\left(-\frac{\omega^2}{2B^2}\right),
\end{equation}
which corresponds to an infinite-dimensional hypercubic lattice.
In this case the GA becomes exact for the energy functional of the
Gutzwiller wave function.
Due to the occurrence of the Brinkman-Rice transition at half filling
$F_0^a$ saturates at a value $F_0^a>-1$ for $U>1$. Thus, in
this particular case, there is no second-order paramagnetic-ferromagnetic
phase transition. The condition $F_0^a=-1$ can be fulfilled in a restricted
doping range, i.e., $0 < \delta < 0.418$, and the corresponding instability
line is shown in the inset of Fig.~\ref{figfla}. 
We find complete agreement of our RPA approach with the phase diagram 
determined by a variational approach in Ref.~\onlinecite{FAZEKAS}.

\begin{figure}[tbp]
\includegraphics[width=7cm,clip=true]{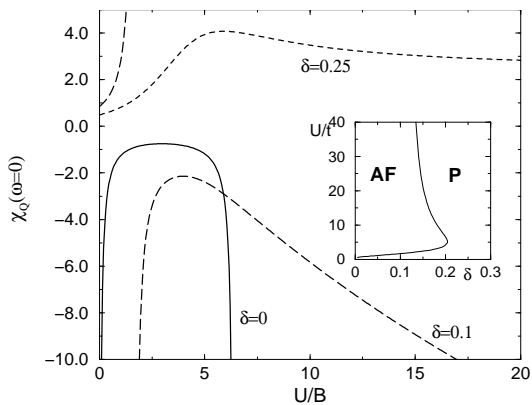}
\caption{RPA static susceptibility $\left[\chi_Q(\omega=0)\right]_{11}$ 
as function of $U/B$ for an infinite
dimensional hypercubic lattice. The inset shows the 
paramagnetic-antiferromagnetic instability line.}
\label{figphaf}
\end{figure}

In order to investigate the instability toward antiferromagnetism,
we study the $\omega=0$ susceptibility at wave vector $Q=(\pi,\pi,\pi, \dots)$.
The inspection of Eq.~(\ref{sus}) reveals that in the case of a 
nearest-neighbor hopping tight-binding band with 
$\varepsilon_{k+q}=-\varepsilon_k$ only the $(1,1)$ matrix element of the 
bare susceptibility is different from zero:
\begin{equation}
\left[\chi_Q^0(\omega=0)\right]_{11}=
\frac{1}{\sqrt{8\pi}B} E_1\left(\frac{1}{2}\frac{\varepsilon_F^2}{B^2}\right),
\end{equation}
where $E_1(x)$ denotes the exponential integral.~\cite{ABST}
The RPA series of Eq.~(\ref{eqrpa}) then leads to
\begin{eqnarray}\label{chiaf}
\left[\chi_Q(\omega=0)\right]_{11}&=& \frac{\left[\chi_Q^0(\omega=0)\right]_{11}}
{1+ N_Q
\left[\chi_Q^0(\omega=0)\right]_{11}}, \\
N_Q&=&e_0\lbrace 2z_0z''-(z')^2\rbrace. \nonumber
\end{eqnarray}
We show the behavior of $\left[\chi_Q^0(\omega=0)\right]_{11}$ for 
various $\delta$ in Fig.~\ref{figphaf}. 

Due to the complete nesting, the bare susceptibility 
$\left[\chi_Q^0(\omega=0)\right]_{11}$ diverges for $\delta=0$.
Hence in this case the singularities of $\left[\chi_Q(\omega=0)\right]_{11}$
are determined by the zeros of the interaction
kernel $N_Q$ which naturally vanishes for $U/B=0$ but
also at the Brinkman-Rice transition where $z_0 \to 0$.
The latter, however, is irrelevant since it occurs in the antiferromagnetic 
phase.
The pole at $U/B=0$ indicates that the
instability toward antiferromagnetism at half filling 
occurs at arbitrarily small interaction also in
infinite dimensions.
For finite $\delta$ the bare magnetic susceptibility is finite and
consequently the pole of $\chi_Q(\omega=0)$ is due to the vanishing of the
RPA denominator in Eq.~(\ref{chiaf}).
It turns out that the static magnetic susceptibility
has exactly one pole in the range $0 \le \delta < 0.117$, 
two poles in the range $0.117 \le \delta < 0.2048$ and no pole
for $\delta \ge 0.2048$.
For completeness, Fig.~\ref{figphaf} also displays $\chi_Q(\omega=0)$ 
for $\delta=0.25$, where there is a small enhancement for those values of $U/B$ 
where the instability occurred for smaller $\delta$. 
The inset of Fig.~\ref{figphaf} shows the antiferromagnetic-paramagnetic 
instability line
constructed from the poles of $\chi_Q(\omega=0)$. Again we find complete
agreement with the variational approach of Ref.~\onlinecite{FAZEKAS}.
Note that one should also determine the first-order
boundaries between the ferromagnetic and antiferromagnetic phases.
Since our intention is limited to a demonstration of the consistency 
of the GA+RPA approach, we refer the reader to Ref.~\onlinecite{FAZEKAS},
where the antiferromagnetic-ferromagnetic phase boundaries have been 
determined by comparing the respective ground-state energies. 

\subsection{Comparison with exact results}\label{section:IIIC}

In the previous subsection we have mainly focused on the static
limit of our RPA approach. This final part is devoted to an
analysis of the magnetic properties of the GA+RPA method, which is
compared to the HF+RPA and the exact results on a $4\times 4$ Hubbard cluster 
with nearest-neighbor hopping.

\subsubsection{Half-filled system}

We start with the half-filled system with
8 spin-up and 8 spin-down particles. The ground state Slater determinant
for the GA and the HF approximation corresponds to a SDW,
which breaks the spin-rotational symmetry of the 
Hamiltonian. As a consequence the transverse magnetic excitations contain 
zero-energy Goldstone modes at wave vector $Q=(\pi,\pi)$.
To avoid numerical instabilities, we have added a small perturbation 
to the Hamiltonian
\begin{equation}\label{vstag}
V=\alpha \sum_i (S_i^z)^2, 
\end{equation}
with $\alpha \sim 10^{-4}t$, which 
shifts the Goldstone modes to small but finite energies ($\sim \alpha$).
In the exact solution an analogue pole appears at small but non-zero
frequency ($\omega/t \approx 0.145$)
due to the finiteness of the cluster. In the thermodynamic
limit long-range order is recovered~\cite{hirsch} and a Goldstone
mode will appear as in the mean-field solution with a weight
related to the order parameter. Here, we are interested in the
finite-frequency behavior and, therefore, we exclude the exact and
approximate ``Goldstone-like''
poles from the comparison and restrict ourself to the 
finite-frequency (triplet) excitations, which, for the chosen value
of $\alpha$, do not sensitively depend on the anisotropy field 
Eq.~(\ref{vstag}).

\begin{figure}[tbp]
\includegraphics[width=8cm,clip=true]{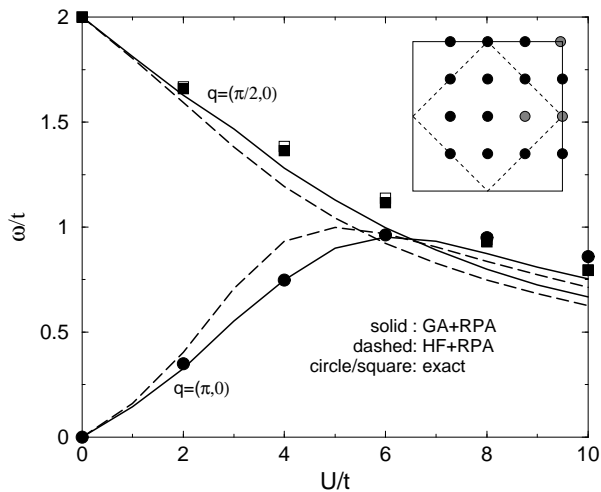}
\caption{
Magnetic excitations at $q=(\pi/2,0)$ and $q=(\pi,0)$ as a function of $U/t$ 
for a half-filled $4\times 4$ cluster:
GA+RPA (solid line), HF+RPA (dashed line), and exact diagonalization 
(full circles and full squares). The exact diagonalization results for the
excitation at $q=(\pi/2,\pi)$ are also reported (empty squares).
In the inset: the q-point mesh of the $4\times 4$ cluster and the dashed
square indicates the doubled Brillouin zone. Shaded points indicates the
important wave vectors of the magnetic excitations.}
\label{n16half}
\end{figure}

Fig.~\ref{n16half} shows the magnetic excitation energies as a
function of $U/t$  evaluated within the GA+RPA, the HF+RPA
and the exact diagonalization. 
Note that the $4\times4$ system has a further accidental 
symmetry, which causes degeneracy between the
$q=(\pi/2,\pi/2)$ and $q=(\pi,0)$ excitations.  
Furthermore, the SDW ground state of the GA and HF solution
leads to the doubling of the Brillouin zone (see inset of Fig.~\ref{n16half})
so that, besides the antiferromagnetic wave vector $Q$,  
only $q=(\pi/2,0)$ and $q=(\pi,0)$ correspond to independent excitations.
On the other hand, on the $4\times 4$ lattice, we have that
the exact energies at $q=(\pi/2,0)$ and $q=(\pi/2,\pi)$ are slightly different.

The small-U behavior of the lowest excitation energy in Fig.~\ref{n16half}
can be well understood from the SDW picture. Within this approximation,
the band structure in the reduced Brillouin zone is given by 
$E_q =\pm \sqrt{\varepsilon_q^2+\Delta^2}$, with $\varepsilon_q=
-2t[\cos(q_x)+\cos(q_y)]$ and $\Delta$ denotes the SDW gap.
Since we study a half-filled system, all states with $E_q < 0$ are
occupied. 
Consider first the $q=(\pi,0)$ excitation which can be attributed 
to a spin-flip transition from $q_1=(-\pi/2,\pm\pi/2)$ to 
$q_2=(\pi/2,\pm\pi/2)$ 
so that the excitation energy is given by $\omega=E_{q_1}-E_{q_2}=2\Delta$.
The SDW gap in the HF approximation is related to
the on-site magnetization $\Delta^{HF}=2 U |S_z|$, whereas within
the KR formulation of the GA it is
determined by the difference in the local spin-dependent Lagrange
multipliers $\Delta^{GA}=\lambda_{\uparrow}-\lambda_{\downarrow}$.
Since in the limit $U\to 0$ the GA reduces to the HF approximation,
both excitation energies coincide in this regime and also agree
with the exact result. On the other hand, for $U/t \gtrsim 1$, where RPA
corrections become important, it can be seen from Fig.~\ref{n16half}
that the GA+RPA is in much better agreement with exact diagonalization
than the corresponding HF+RPA result.
As a consequence, the GA+RPA gives a quite accurate description of the 
crossover (at $U/t \approx 6$) from the SDW regime, 
where a gap proportional to $U$ opens 
along the Fermi surface, to the Heisenberg regime, where there are low-energy 
magnetic excitations with energy scale $t^2/U$. 

For the higher energy triplet excitation at $q=(\pi/2,0)$, the GA+RPA yields
energies which are slightly lower than the exact result. However,
whereas the discrepancy for the GA+RPA 
at $U/t=6$ is around $10\%$, the HF+RPA deviates by almost $20\%$
from the exact diagonalization result.
\begin{figure}[tbp]
\includegraphics[width=8cm,clip=true]{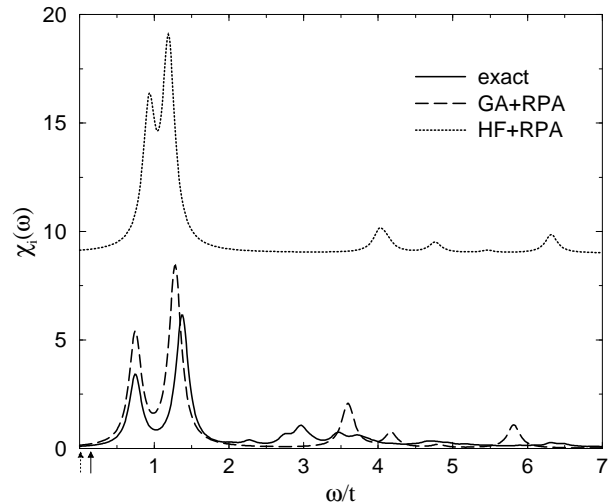}
\caption{Local magnetic susceptibility $\chi(\omega)$ for the half-filled 
$4\times4$ cluster
calculated within exact diagonalization, GA+RPA and HF+RPA for $U/t=4$. 
The HF+RPA curve has been shifted for convenience. The two arrows indicate
the energy of the lowest $Q=(\pi,\pi)$ excitation at $\omega/t \approx
0.145$ (exact diagonalization) and $\omega/t=0$ (RPA Goldstone mode).}
\label{u4h16sus}
\end{figure}

In order to have information on the accuracy of the GA+RPA for finite
frequencies, we report in Fig.~\ref{u4h16sus} the local magnetic susceptibility
\begin{equation}\label{chi}
\chi(\omega)=\sum_q\sum_{m>0}|\langle \Psi_m
|S_q^+|\Psi_0\rangle|^2\delta(\omega-(E_m-E_0)),
\end{equation}
for the GA+RPA and the HF+RPA approximations and the exact diagonalization for
$U/t=4$. 
The $\delta$-functions in Eq.~(\ref{chi}) have been replaced by Lorentzians
with width $0.1t$.

\begin{figure}[tbp]
\psfrag{inset6}{\bf $\int d\omega \omega \chi(\omega)$}
\includegraphics[width=8cm,clip=true]{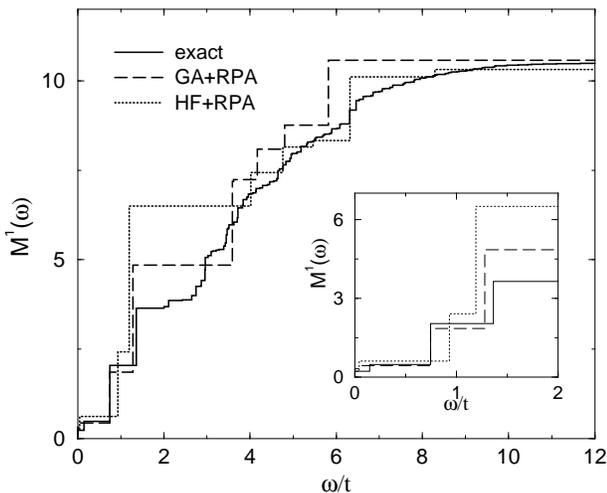}
\caption{Cumulative sum of the first moment of $\chi(\omega)$
for the exact result, GA+RPA, and HF+RPA. Data are for a half-filled 
$4\times4$ cluster and $U/t=4$. Inset: a detail of the low-energy part.}
\label{u4h16cum}
\end{figure}

The two lowest-energy excitations are quite accurate
within the GA+RPA approach except for a moderate   
overestimation of the intensity which becomes much worse in  
HF+RPA. Interestingly, also the high-energy spin fluctuations are
in the correct frequency range. 

It should be noted that the spectral weight is constrained by the 
following sum rule:  
\begin{equation}\label{SR}
\int_0^\infty\!\mbox{d}\omega \omega \chi(\omega) = 
-\frac{1}{2}\langle T \rangle_{GA},
\end{equation}
where $\langle T \rangle_{GA}$ is the average value of the kinetic energy
in the GA.
The sum rule Eq.~(\ref{SR})
relates the RPA correlation function $\chi(\omega)$ to the kinetic
energy computed within the GA. A similar sum rule is valid in HF+RPA
with the kinetic energy computed in HF. 
In Ref.~\onlinecite{GOETZ1}, we have already demonstrated
that the GA kinetic energy is in remarkable agreement with the exact result
over a large doping range which in the present context gives additional 
support to the GA+RPA approach also in the magnetic sector.
On the other hand, the HF approximation is of inferior quality in describing  
excitation energies and the total kinetic energy.
Therefore it is not surprising that also spectral weights 
perform much worse than in the GA+RPA approach. 

Finally, Fig.~\ref{u4h16cum} displays the frequency evolution of the 
first moment of $\chi(\omega)$:
\begin{equation}
M^1(\omega)=\int_0^\omega\!\!d\tilde{\omega}\, \tilde{\omega} \chi(\tilde{\omega}),
\end{equation}
for the same parameters as in Fig.~\ref{u4h16sus}.
Note that $M^1(\omega)$ contains the contribution
from the lowest $Q=(\pi,\pi)$ excitations (i.e., the Goldstone modes
in the mean-field plus RPA), which appear as the offsets at small
energies.
From the inset of Fig.~\ref{u4h16cum}, it can be seen that, especially at 
low frequencies, the GA+RPA approach provides a much better
approximation for the exact $M^1(\omega)$ than HF+RPA,
both with respect to the poles and intensities.
Both the HF+RPA and the GA+RPA approximate the
incoherent part of the exact spectrum (i.e., for $\omega/t>2$) by a rather 
small set of excitations. However, the corresponding step-like evolution 
of the first moment of $\chi(\omega)$ is quite close to the exact result.

\subsubsection{Doped system}

\begin{figure}[tbp]
\includegraphics[width=8cm,clip=true]{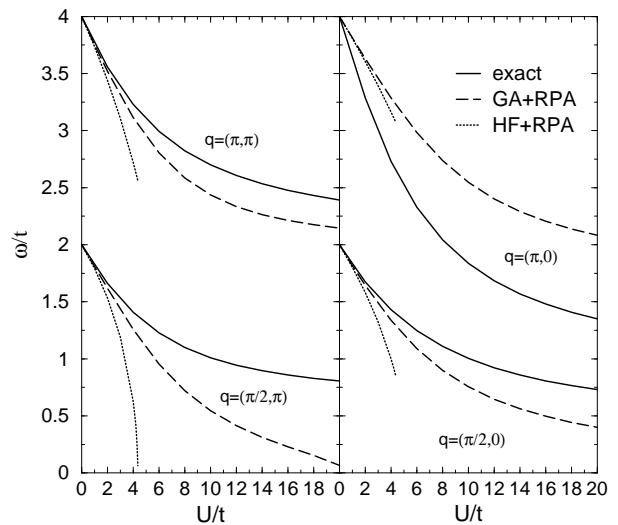}
\caption{
Magnetic excitations for wave vectors $q=(\pi,\pi)$ and $q=(\pi/2,\pi)$
(left panel) and for $q=(\pi,0)$, $q=(\pi/2,0)$ (right panel) as a function of
$U/t$ for the exact diagonalization (solid line), GA+RPA (dashed line),
and HF+RPA (dotted line). The HF+RPA are shown for $U<U_{crit}$.}
\label{n10half}
\end{figure}

We finally investigate a $4\times 4$ cluster with 5 spin-up and 5 spin-down 
particles, corresponding to a closed-shell configuration.
The HF solution undergoes a magnetic instability
with wave vector $q=(\pi/2,\pi)$ at $U_{crit}/t \approx 4.365$,
marked by the softening of the corresponding excitation at this
critical value. For $U>U_{crit}$ the HF+RPA  spectrum 
has a Goldstone mode and in general the performance is very poor 
consistently with the fact that a broken symmetry state is not
expected even in the thermodynamic limit, therefore, we restrict to 
the more physical paramagnetic solution.
In the GA case the paramagnetic solution can be stabilized
for much larger values of $U/t$ providing a reasonable starting point
in a broader parameter range  just as in the two-site case.
The GA+RPA approach 
captures the behavior of the exact solution (namely
the softening of triplet excitations) at least in a qualitative way, although
quantitative deviations increase with increasing $U/t$ (see Fig.~\ref{n10half}).

\begin{figure}[tbp]
\psfrag{inset6}{\bf $M^1(\omega)$}
\includegraphics[width=8cm,clip=true]{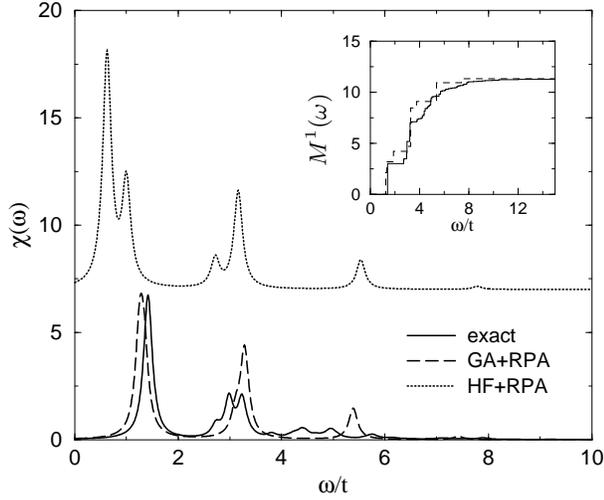}
\caption{Local magnetic susceptibility $\chi(\omega)$ for the $4\times4$ 
cluster with 10 particles
calculated within exact diagonalization, GA+RPA and HF+RPA for $U/t=4$. 
The HF+RPA curve has been shifted for convenience. 
The inset shows the cumulative sum of the first moment of $\chi(\omega)$
for the exact result and the GA+RPA.}
\label{u4n10sus}
\end{figure}

In Fig.~\ref{u4n10sus}, we compare the local susceptibility of 
the HF+RPA, the GA+RPA, and the exact diagonalization for $U/t=4$, i.e.,
for values of $U/t$ smaller than the magnetic instability.
The GA+RPA not only gives a rather good estimate
to the lowest excitation energy but in addition provides a good approximation
for the corresponding intensity. Note that, since for the given value of
$U/t$ the HF solution is already close to the 
$q=(\pi/2,\pi)$ instability, we observe
a strong softening of the lowest energy excitation 
resulting in a significantly  enhanced oscillator strength.

Finally, we have also evaluated the cumulative
integral of the first moment of $\chi(\omega)$ for the GA+RPA and
exact diagonalization, which is shown in the inset of
Fig.~\ref{u4n10sus}. Due to the sum rule Eq.~(\ref{SR}) and 
the excellent kinetic energy approximation of the GA, the integrated
weight of the GA+RPA and the exact diagonalization are in excellent agreement. 
Moreover, we again observe that the GA+RPA provides a rather good
step-like approximation to the exact evolution of the spectral
weight as a function of the frequency.

\section{Conclusion}\label{section:IV}

In this paper, we have presented a detailed investigation of the quality of 
the GA+RPA approach for the computation of the magnetic excitations. 
The present computation is complementary to the previous computation in the
charge sector.~\cite{GOETZ1,GOETZ2}  
An unexpected outcome of the present work is a further justification
of the antiadiabatic assumption for the time-evolution of the double
occupancy, which was needed in the charge but not in the magnetic channel.
The fact that the two calculations for the spin and the charge sector
give the correct degeneracy of the excitation spectrum for a spin-rotational 
invariant system (the two-site Hubbard model) clearly indicates
that such an assumption was indeed correct, i.e., other
possibilities like to keep the double occupancy fixed at the
stationary value (rather than to follow the time evolution of the
density matrix) would had lead to an unphysical breaking of 
spin-rotational symmetry.
  
The present formalism is based on a Gutzwiller-type
energy functional, which can be either obtained from the 
spin-rotational invariant KR slave-boson scheme~\cite{WH} or alternatively 
from the standard GA with spin-rotated Slater determinants.~\cite{MICNAS} 
In our approach, due to the fact that all bosonic fields
have been already eliminated from the saddle-point energy functional,
it turns out that the evaluation of RPA fluctuations
around the GA solution is significantly simplified.
In the present paper, we have restricted the calculations
of the magnetic excitations to small Hubbard clusters in order 
to compare with exact diagonalization results.
The better performance of GA+RPA with respect to 
HF+RPA has been demonstrated for both excitation energies and the
corresponding intensities.
However, compared to numerical methods~\cite{LANCZOS}
our approach can be pushed to much larger systems.
In particular, it is suitable for the evaluation of magnetic
excitations around inhomogeneous solutions of Hubbard-type models,
where it is constrained to the same size limitations  
than the unrestricted HF+RPA approach. This is interesting in 
connection with the magnetic susceptibility in nickelates and 
high-T$_c$ cuprates, which are both characterized 
by the presence of strong electronic 
correlations and inhomogeneous charge distributions in some part of
the phase diagram. Work in this direction is in progress. 

\acknowledgments
G.S. and P.R. gratefully acknowledge financial support from the Deutsche 
Forschungsgemeinschaft, and F.B. the support from INFM.

\appendix

\section{Derivatives of the hopping factor}\label{APA}

The derivatives appearing in Eqs.~(\ref{z1}) and~(\ref{z2}) are given by:
\begin{widetext}
\begin{eqnarray}
\frac{\partial z_{i,\sigma,-\sigma}}{\partial \rho_{ii}^{-\sigma,\sigma}}
&=& \frac{1}{2}\frac{z_{i\uparrow\uparrow}^0 - z_{i\downarrow\downarrow}^0}{S_i^z}, \\
\frac{\partial^2 z_{i,\sigma \sigma}}
{\partial \rho_{ii}^{\sigma, -\sigma}\partial \rho_{ii}^{-\sigma, \sigma}}
&=& \frac{\sigma}{2 S_i^z}\left \lbrace
-\frac{z_{i\uparrow\uparrow}^0 - z_{i\downarrow\downarrow}^0}{S_i^z}
+{\sqrt{\rho_i^{\sigma\sigma}
(1-\rho_i^{\sigma\sigma})}}\left[\frac{\sqrt{1-\rho_{ii}+D_i}}{\sqrt{
\rho_i^{\sigma\sigma}-D_i}}-\frac{\sqrt{D_i}}{\sqrt{\rho_i^{-\sigma-\sigma}}}
\right]
-\frac{1-2\rho_i^{\sigma\sigma}}{\rho_i^{\sigma\sigma}
(1-\rho_i^{\sigma\sigma})}z_{i,\sigma\sigma}^0\right \rbrace.
\end{eqnarray}
In case of a homogeneous, paramagnetic saddle point these
expressions simplify to
\begin{eqnarray}
\frac{\partial z_{i,\sigma,-\sigma}}{\partial \rho_{ii}^{-\sigma,\sigma}}
\equiv z' &=& \frac{2\delta}{1-\delta^2}\left(\frac{1}{z_0} - z_0\right), \\
\frac{\partial^2 z_{i,\sigma \sigma}}
{\partial \rho_{ii}^{\sigma, -\sigma}\partial \rho_{ii}^{-\sigma, \sigma}}
\equiv z'' &=& \frac{2z_0}{(1-\delta^2)^2}\left\lbrace 1-2\delta^2\left(\frac{1}{z_0^2}
-1\right)\right\rbrace -\frac{1}{2}\frac{z_0}{(1-\delta-2D)^2},
\end{eqnarray}
\end{widetext}
where $z_0$ denote the elements of the z-matrix defined in Eq.~(\ref{z0}).

\section{Derivation of the RPA matrices}\label{APB}

In order to give explicit expressions for the RPA matrices $A$ and $B$
as defined in Eq.~(\ref{phe}) one has first to diagonalize
the Gutzwiller Hamiltonian Eq.~(\ref{eq:hgw}) via
\begin{equation}\label{TRAFO1}
c_{i,\sigma}=\sum_{\nu}\Phi_{i}(\nu,\sigma)a_{\nu,\sigma},
\end{equation}
where $\nu$ refers to either particle (p) or hole (h) states.
Inserting this transformation in the expansion Eq.~(\ref{espin})
leads to the following expressions for $A$ and $B$:
\begin{eqnarray} \label{A}
A_{ph,p'h'}^{\sigma\sigma'} &=&  (\varepsilon_{p\sigma}
-  \varepsilon_{h,-\sigma})\delta_{pp'}\delta_{hh'}\delta_{\sigma\sigma'}
\nonumber \\
&+& \delta_{\sigma \sigma'}
\sum_{ij} N_{ij} \Phi_i(p\sigma) \Phi_i(h,{-\sigma})
 \Phi_j(p'{\sigma}) \Phi_j(h',{-\sigma})
\nonumber \\
&+& R_{ph,p'h'}^{\sigma\sigma'}+R_{p'h',p h}^{-\sigma\sigma'},
\nonumber \\
B_{ph,p'h'}^{\sigma\sigma'}&=& \delta_{\sigma,- \sigma'} \sum_{ij}
N_{ij} \Phi_i(p \sigma) \Phi_i(h, -\sigma)
\nonumber \\
&\times& \Phi_j(p',-\sigma) \Phi_j(h'  \sigma))+
T_{ph,p'h'}^{\sigma\sigma'}+T_{p'h',ph}^{\sigma\sigma'},
\end{eqnarray}
where
\begin{eqnarray} \label{N}
N_{ij}&=&2 \delta_{ij} \sum_{n,\sigma} t_{nj} z^0_{n,\sigma \sigma}
\rho^0_{nj,\sigma \sigma}
\frac{\partial^2 z_{j,\sigma \sigma}}
{\partial (\rho_{jj,\uparrow \downarrow}) \partial (\rho_{jj,\downarrow \uparrow})} \nonumber \\
&+& t_{ij} \frac{\partial z_{i,\uparrow   \downarrow}}{\partial \rho_{i,\downarrow \uparrow  }}
      \frac{\partial z_{j,\downarrow \uparrow  }}{\partial \rho_{j,\uparrow   \downarrow}}
(\rho^0_{ij,\uparrow \uparrow}+\rho^0_{ij,\downarrow \downarrow}) (1-\delta_{ij}), \nonumber \\
T_{ph,p'h'}^{\sigma\sigma'} &=&\delta_{\sigma \sigma'}
\delta_{\sigma \uparrow} \sum_{ij} t_{ij} \frac{\partial
z_{j,\downarrow \uparrow}}{\partial \rho_{jj,\uparrow \downarrow}}
\Phi_{j}(p \uparrow) \Phi_j(h \downarrow)
\nonumber \\
&\times& \left( z^0_{i,\uparrow \uparrow    } \Phi_i(p^\prime
\uparrow) \Phi_j(h^\prime \downarrow) + z^0_{i,\downarrow
\downarrow} \Phi_j(p^\prime \uparrow) \Phi_i(h^\prime \downarrow)
\right)
\nonumber \\
&+&\delta_{\sigma, \sigma'} \delta_{\sigma \downarrow} \sum_{ij}
t_{ij} \frac{\partial z_{j,\uparrow \downarrow}}{\partial
\rho_{jj,\downarrow \uparrow}} \Phi_{j}(p \downarrow) \Phi_j(h
\uparrow)
\nonumber \\
&\times& \left( z^0_{i,\uparrow \uparrow    } \Phi_j(p^\prime
\downarrow) \Phi_i(h^\prime \uparrow) + z^0_{i,\downarrow
\downarrow} \Phi_i(p^\prime \downarrow) \Phi_j(h^\prime \uparrow)
\right),
\nonumber \\
R_{ph,p'h'}^{\sigma\sigma'} &=&\delta_{\sigma,- \sigma'}
\delta_{\sigma \uparrow} \sum_{ij} t_{ij} \frac{\partial
z_{j,\downarrow \uparrow}}{\partial \rho_{jj,\uparrow \downarrow}}
\Phi_{j}(p \uparrow) \Phi_j(h \downarrow)
\nonumber \\
&\times& \left( z^0_{i,\uparrow \uparrow    } \Phi_i(h^\prime
\uparrow) \Phi_j(p^\prime \downarrow) + z^0_{i,\downarrow
\downarrow} \Phi_j(h^\prime \uparrow) \Phi_i(p^\prime \downarrow)
\right)
\nonumber \\
&+&\delta_{\sigma, -\sigma'} \delta_{\sigma \downarrow} \sum_{ij}
t_{ij} \frac{\partial z_{j,\uparrow \downarrow}}{\partial
\rho_{jj,\downarrow \uparrow}} \Phi_{j}(p \downarrow) \Phi_j(h
\uparrow)
\nonumber \\
&\times& \left( z^0_{i,\uparrow \uparrow    } \Phi_i(p^\prime
\uparrow) \Phi_j(h^\prime \downarrow) + z^0_{i,\downarrow
\downarrow} \Phi_j(p^\prime \uparrow) \Phi_i(h^\prime \downarrow)
\right).
\nonumber \\
\end{eqnarray}


\begin{thebibliography}{9}
\bibitem{GUTZ1} M.C. Gutzwiller, Phys. Rev. Lett. {\bf 10}, 159 (1963).
\bibitem{GUTZ2} M.C. Gutzwiller, Phys. Rev. {\bf 134}, A 923 (1964);
    {\it ibid.} {\bf 137}, A 1726 (1965).
\bibitem{METZNER} W. Metzner and D. Vollhardt, Phys. Rev. Lett. {\bf 59},
    121 (1987); Phys. Rev. B {\bf 37}, 7382 (1988).
\bibitem{VOLLHARDT} D. Vollhardt, Rev. Mod. Phys. {\bf 56}, 99 (1984).
\bibitem{GOETZ} G. Seibold, C. Castellani, C. Di Castro, and M. Grilli, Phys.
    Rev. B {\bf 58}, 13506 (1998).
\bibitem{LOR} J. Lorenzana and G. Seibold, Phys. Rev. Lett. {\bf 89},
    136401, (2002).
\bibitem{GEBHARD} F. Gebhard, Phys. Rev. B {\bf 41}, 9452 (1990).
\bibitem{michele} C. Attaccalite and M. Fabrizio, Phys. Rev. B {\bf 68}, 
    155117 (2003).
\bibitem{KR} G. Kotliar and A.E. Ruckenstein, Phys. Rev. Lett. {\bf 57},
    1362 (1986).
\bibitem{ARRIGONI} E. Arrigoni and G. C. Strinati, Phys. Rev. Lett. {\bf 71},
    3178 (1993); Phys. Rev. B {\bf 52}, 2428 (1995).
\bibitem{LAVAGNA} M. Lavagna, Phys. Rev. B {\bf 41}, 142 (1990).
\bibitem{LI} J.W. Rasul and T. Li, J. Phys. C {\bf 21}, 5119 (1988).
\bibitem{rai93} R. Raimondi and C. Castellani, Phys. Rev. B {\bf 48}, 
    R11453 (1993).
\bibitem{rai95} R. Raimondi, Phys. Rev. B {\bf 51}, 10154 (1995).
\bibitem{raipc} R. Raimondi, private communication.
\bibitem{GOETZ1} G. Seibold and J. Lorenzana, Phys. Rev. Lett. {\bf 86},
    2605 (2001).
\bibitem{GOETZ2} G. Seibold, F. Becca, and J. Lorenzana, Phys. Rev. B {\bf 67},
    085108 (2003).
\bibitem{THOULESS} D.J. Thouless, Nucl. Phys. {\bf 22}, 78 (1961).
\bibitem{YON} K. Yonemitsu, I. Batisti\'{c}, and A. R. Bishop,
    Phys. Rev. B {\bf 44}, 2652 (1991); 
    K. Yonemitsu, A. R. Bishop, and J. Lorenzana, Phys. Rev. B {\bf 47},
    12059 (1993).
\bibitem{lor03} J. Lorenzana and G. Seibold, Phys. Rev. Lett. {\bf 90}, 
    066404 (2003).
\bibitem{BM} J. B\"unemann, J. Phys.: Condens. Matter {\bf 13}, 5321 (2001).
\bibitem{FRESARD} W. Zimmermann, R. Fr\'{e}sard, and P. W\"olfle, Phys. Rev.
    B {\bf 56}, 10097 (1997).
\bibitem{WH} T. Li, P. W\"olfle, and P. J. Hirschfeld, Phys. Rev. B {\bf 40},
    6817 (1989); R. Fr\'{e}sard and P. W\"olfle, Int. J. Mod. Phys.
    B {\bf 6}, 237 (1992).
\bibitem{MICNAS} M. Bak and R. Micnas, J. Phys.: Condens. Matter {\bf 10},
    9029 (1998).
\bibitem{SSH} G. Seibold, E. Sigmund, and V. Hizhnyakov, Phys. Rev. B {\bf 57},
    6937 (1998).
\bibitem{goe98} G. Seibold, Phys. Rev. B {\bf 58}, 15520 (1998).
\bibitem{RING} P. Ring and P. Schuck, {\it The nuclear many-body problem}
    Springer-Verlag, New York, 1980.
\bibitem{BLAIZOT} J. Blaizot, G. Ripka {\it Quantum theory of finite syatems},
    MIT Press, 1986. 
\bibitem{FAZEKAS} P. Fazekas, B. Menge, and E. M\"uller-Hartmann,
    Z. Phys. B {\bf 78}, 69 (1990).
\bibitem{ABST} M. Abramowitz and I. A. Stegun (eds.), 
    {\it Handbook of Mathematical Functions} (Dover Publications 1965).
\bibitem{hirsch} J.E. Hirsch, Phys. Rev. B {\bf 31}, 4403 (1985);
    J.E. Hirsch and S. Tang, Phys. Rev. Lett. {\bf 56}, 273 (1989);
    S.R. White, D.J. Scalapino, R.L. Sugar, E.Y. Loh, J.E. Gubernatis,
    and R.T. Scalettar, Phys. Rev. B {\bf 40}, 506 (1989).
\bibitem{LANCZOS} E. Dagotto, Rev. Mod. Phys. {\bf 66}, 763 (1994).
\end{thebibliography}
\end{document}